\documentclass[twocolumn,showpacs,preprintnumbers,amsmath,amssymb]{revtex4}

\usepackage{graphicx}
\usepackage{dcolumn}
\usepackage{bm}
\usepackage{amsmath}
\usepackage{amssymb}
\usepackage{mathrsfs}

\newcommand{\p}{\partial}
\newcommand{\dd}{{\rm d}}
\newcommand{\e}{\epsilon}

\newcommand{\sd}{Schr\"{o}dinger }
\newcommand{\pt}{P\"{o}schl-Teller }
\newcommand{\sech}{{\rm sech} }

\newcommand{\A}{\mathfrak{A}}
\newcommand{\B}{\mathfrak{B}}
\newcommand{\C}{\mathfrak{C}}
\newcommand{\G}{\mathcal{G}}
\newcommand{\h}{\mathscr{H}}

\newcommand{\Df}{{\rm Diff}}
\newcommand{\la}{\mathfrak{L}}
\newcommand{\Dw}{{\mathscr{D}_\omega}}
\newcommand{\Ds}{{\mathscr{D}_\infty}}

\newcommand{\M}{\mathbf{M}}
\newcommand{\R}{\mathscr{R}}
\newcommand{\Exp}{{\rm Exp}}
\newcommand{\ad}{{\rm ad}}

\newtheorem{theorem}{Theorem}[section]
\newtheorem{lemma}{Lemma}[section]

\newtheorem{definition}{Definition}[section]

\newtheorem{proposition}{Proposition}[section]
\begin{document}


\title{Smooth Controllability of Infinite Dimensional Quantum Mechanical Systems}

\author{Re-Bing Wu}
 \altaffiliation{Department of Chemistry, Princeton
University, Princeton, NJ 08544, USA}
\email{rewu@Princeton.EDU}

\author{Tzyh-Jong Tarn}%
\altaffiliation{Department of Electrical and Systems Engineering,
Washington University, St. Louis, MO 63130, USA}
\email{tarn@wuauto.wustl.edu}

\author{Chun-Wen Li}
\altaffiliation{Department of Automation, Tsinghua University,
Beijing, 100084, P. R. China} \email{lcw@mail.tsinghua.edu.cn}

\date{\today}

\begin{abstract}
Manipulation of infinite dimensional quantum systems is important
to controlling complex quantum dynamics with many practical
physical and chemical backgrounds. In this paper, a general
investigation is casted to the controllability problem of quantum
systems evolving on infinite dimensional manifolds. Recognizing
that such problems are related with infinite dimensional
controllability algebras, we introduce an algebraic mathematical
framework to describe quantum control systems possessing such
controllability algebras. Then we present the concept of smooth
controllability on infinite dimensional manifolds, and draw the
main result on approximate strong smooth controllability. This is
a nontrivial extension of the existing controllability results
based on the analysis over finite dimensional vector spaces to
analysis over infinite dimensional manifolds. It also opens up
many interesting problems for future studies.
\end{abstract}

\pacs{03.67.-a,02.20.Qs,02.20.-a  }
\keywords{Quantum control, controllability, Lie algebra}
\maketitle

\section{Introduction}
The laboratory successes in coherent control of quantum dynamics
in late 1990s have dawned the dream of controlling quantum
phenomena. Nowadays, advances in laser technologies and system
control theory \cite{tarn2,tarn5,rabitz2,rabitz3} have
demonstrated the abilities of effectively manipulating microscopic
systems both theoretically and experimentally. Owing to the unique
quantum coherence, quantum control has manifested incredible
novelties in contrast to the corresponding classical control
schemes \cite{rabitz6}. With new perspectives obtained from these
achievements, more ambitious goals are being pursued to control
complex quantum dynamics subject to issues such as large molecules
\cite{herek}, entanglements in quantum networks \cite{blais,mao},
and decoherence in open quantum systems \cite{lloyd2,jing}.
Moreover, the developments of quantum measurements, which are in
general weak and continuous in time, make it possible to implement
feedback techniques to enhance the ability and robustness of
control of non-unitary noisy evolutions of quantum states
\cite{mao,wiseman2}.

In this paper, we are concerned with quantum control of infinite
dimensional systems, especially those that contain continuous
spectra, which are fundamental in many practical backgrounds. For
example, as a long-standing problem in controlling ultrafast
molecular dynamics, the attempts to break chemical bonds naturally
fall under transitions of molecular states between discrete and
continuous spectra \cite{alhassid6,rice3,shapiro1}. More recently,
new motivations are generated by the development of continuous
quantum computer \cite{lloyd4,contqc} that processes quantum
information encoded in continuous spectra. Serious theoretical
studies have proved that they might be more sufficient in some
tasks in comparison to their discrete counterparts.

A large class of infinite dimensional quantum systems with
discrete spectra can be well approximated to possess a finite
number of levels under weak control field interactions that induce
transitions between these levels. Mathematical treatments to such
systems can then be largely simplified by linear approximations
\cite{zuazua,weaver}, perturbation theory or adiabatic
approximations \cite{shapiro1}. However, in intense-field
circumstances numerous levels have to be taken into account
because they become strongly coupled by multi-photon processes.
For systems with continuous spectra, some of the above
approximations will not be applicable even in weak-field cases due
to the non-local nature of corresponding scattering states. To
exert the use of physical resources in control, we need to put the
study back into the infinite dimensional prototype model of
quantum mechanical systems so that arbitrary spectral types can be
universally handled. To the authors' knowledge, not much
theoretical consideration has been received from this perspective.
The first studies date back to the work of Huang, Tarn and Clark
\cite{tarn1,tarn2} in which analytic controllability is
systematically studied based on group representation theory. Their
work was later extended to time-dependent systems by Lan, Tarn,
Chi and Clark \cite{lan1}. These results embrace both cases of
discrete and continuous spectra \cite{tarn5,lan1}, however, are
still restricted to systems with finite dimensional
controllability algebras. Some other specific discussions can be
found in \cite{brockett1,karwowski,lloyd1}.

In this paper, the study of infinite dimensional quantum systems
will be embedded in an algebraic framework that can deal with
infinite dimensional controllability Lie algebras. This framework
also allows the use of strong fields and unbounded Hamiltonians to
enforce control over the quantum states. On such a basis, we
present new concepts of smooth controllability and extend the
existing results to more general systems on infinite dimensional
manifolds. The organization of the paper is as follows: in section
II, we summarize the existing results and present the algebraic
framework accompanied with preliminary examples; in section III,
we give the notions of smooth controllabilities based on the
smooth domain; in section IV, the main result is given and proved
on approximate strong smooth controllability; in section V, we
discuss two typical examples. Finally, in section VI, we draw
conclusions and give our perspectives.

\section{Problem Formulation}\label{problem}

Generally, quantum control systems can be written in the form of
the following \sd equation:
\begin{equation}\label{abc'}
  i\hbar{\p\over\p{t}}\psi(t)=\left(H_0'+H_I'\right)\psi(t),\,\psi(0)=\psi_0,
\end{equation}
where the quantum state $\psi(t)$ evolves in a separable Hilbert
space $\h$. The free (unperturbed or internal) Hamiltonian $H_0'$
of the quantum system is a Hermitian operator on $\h$. The $H_I'$
refers to the interaction (control) Hamiltonian that is used to
affect the quantum dynamics. In most situations, the interaction
Hamiltonian can be decomposed into a sum
$H_I'=\sum_{j=1}^m{u_j(t)H_j'}$, where the $u_i$'s are the
controls that represent some classical fields interacting with the
system through the Hamiltonians in the summation. To simplify the
notations, we rewrite (\ref{abc'}) with skew-Hermitian operators
$H_i=(i\hbar)^{-1}H_i'$, which leads to the following quantum
control system:
\begin{equation}\label{abc}
 {\p\over\p{t}}\psi(t)=\left[H_0+\sum_{j=1}^m{u_j(t)H_j}\right]\psi(t),\,\psi(0)=\psi_0.
\end{equation}

Throughout this paper, we always assume that the controls are
piecewise constant functions of time. A quantum state $\psi'$ is
said to be reachable from $\psi$ if there exists a time instant
$T$ and some admissible (i.e. piecewise constant in this paper)
controls over $[0,T]$ that steers the system from $\psi$ to
$\psi'$. Denote $\R(\psi)$ as the reachable set of all states that
are reachable from $\psi$, and $\R_t(\psi)$ as the reachable set
of all states that are reachable from $\psi$ at a specified time
$t>0$.

The controllability issue concerns itself with the problem of
whether or not the reachable set of the initial state can fill up
a prescribed set of quantum states. In particular, let the
manifold $\M$ be the closure of the set of states \cite{brockett1}
\begin{equation}\label{m}
\{e^{s_kH_{\alpha_k}}\cdots e^{s_1H_{\alpha_1}}\psi_0:\,\,
s_k\in\mathbb{R};\,\alpha_k=0,1,\cdots,m;\,k\in\mathbb{N}\},
\end{equation}
where $\psi_0\in\h$ is the initial state, the reachable set of
$\psi_0$ is obviously contained in $\M$ because all possible
control actions on the system are involved. In the rest parts of
this paper, we will concentrate the studies on controllability
properties on this manifold.

Controllability problem is of great interests mainly in two
aspects. Firstly, since the internal Hamiltonian $H_0$ may give
rise to unwanted complex dynamics (e.g. chaos), the
controllability characterizes the ability of the control system to
fight against such complexities. Secondly, the Yes-or-No answer of
controllability provides important information in many practical
problems, e.g. possibility of $100\%$ ratio of preferred products
in chemical reactions, or universality of quantum computation
realized by some physical structures.

\subsection{Existing Results}

Denote $\A=\{H_0,H_1,\cdots,H_m\}_{LA}$ the controllability
algebra, where the subscript "$LA$" denotes the Lie algebra
generated by operators in the curly bracket. For quantum control
systems with a finite number of levels, the basic result is
summarized as follows \cite{rabitz1}:

\begin{theorem}\label{fc}
Suppose $\h$ is a $N$-dimensional Hilbert space of quantum states
and $S_\h$ is the unit sphere in $\h$. The system is controllable
over $S_\h$, if the controllability algebra $\A=su(N)$ or
$\A=u(N)$. For homogeneous quantum control systems, i.e. $H_0=0$,
the condition is also a necessary condition.
\end{theorem}

Systems with an infinite number of levels are much more
complicated because the Hamiltonians may bring severe domain
constraints. Providing that some of them are unbounded operators
\cite{barut}, the system evolution will have to be restricted in a
proper subset of quantum states in $\h$, on which the Hamiltonians
are well-defined, invariant and the state-evolution can be
expressed globally in exponential form. For system with a finite
dimensional controllability Lie algebra
$\A=\{H_0,H_1,\cdots,H_m\}_{LA}$, Huang, Tarn and Clark
\cite{tarn2} have suggested the analytic domain:
$$\Dw=\left\{\omega\in\h:\,\sum_{n=0}^\infty{\sum_
{1\leq{i_1,\cdots,i_n}\leq{m}}\frac{\|H_{i_1}\cdots{H_{i_n}}\omega\|s^n}{n\,!}}<\infty\right\}$$
\noindent as a candidate. The existence of a dense analytic domain
in $\h$ (with respect to the Hilbert space topology) and
corresponding group representation are guaranteed by the Nelson's
theorem \cite{nelson}). Based on the analytic domain, the notion
of analytic controllability is as follows:
\begin{definition}{\bf (Analytic Controllability)}
Quantum mechanical control system (\ref{abc}) is said to be
analytically controllable on $\M$ if the reachable set
$\R(\psi)=\M\cap\Dw$ for all $\psi\in\M\cap\Dw$. If the reachable
set $\R_t(\psi)$ equals to $\M\cap\Dw$ for all $\psi\in\M\cap\Dw$
at any time $t>0$, the system is said to be strongly analytically
controllable.
\end{definition}

Huang, Tarn, and Clark (HTC) presented a criterion of strong
analytic controllability \cite{tarn2}:
\begin{theorem}\label{analytic contr.}
Suppose the controllability algebra $\A$ is finite dimensional and
the analytic domain exists. Let the Lie algebra
$\B=\{H_1,\cdots,H_r\}_{LA}$ and
$\C=\{\ad_{H_0}^k\B;\,k=0,1,\cdots\}_{LA}$, where
$\ad_{H_0}^0\B=\B$ and $\ad_{H_0}^{k+1}\B=[H_0,\ad_{H_0}^k\B]$.
The system (\ref{abc}) is strongly analytically controllable if
the following conditions are satisfied:
\begin{enumerate}
    \item $[\,\B,\C\,]\subseteq{\B}$;
    \item For any $\phi\in\M\cap\Dw$, ${\rm dim}\,\C(\phi)={\rm dim}\,\M$.
\end{enumerate}
\end{theorem}

Interested readers are further referred to \cite{lan1} for
extended results on time-dependent quantum control systems. A
common restriction of these results is that at most a finite
dimensional manifold $\M$ in finite or infinite dimensional
Hilbert spaces can be taken into consideration, because the
corresponding controllability Lie algebra $\A$ is finite
dimensional. This limitation is manifested more clearly in the
following HTC No-Go theorem \cite{tarn2}:
\begin{theorem}\label{No-Go} The system (\ref{abc}) is not
strongly analytically controllable on $S_\h$, if the
controllability Lie algebra $\A$ is finite dimensional.
\end{theorem}

The negative assertion of No-Go theorem implies strongly the
necessity to explore quantum systems with infinite dimensional
controllability algebras in order to drive systems over infinite
dimensional manifolds. This is the primary motivation of the study
in this paper.

\subsection{Algebraic Model and Preliminary Examples}
Quantum Hamiltonians in traditional models are usually expressed
as combinations of kinetic and potential energies according to the
Hamiltonian formulation of mechanics. Specialized scalar or vector
potentials are applied to affect the quantum dynamics. For
example, the dipole interaction of an electrical field with atoms
or molecules are widely used for quantum control \cite{rabitz2}.
Such expressions are physically explicit, however, not convenient
for calculation in structural analysis of quantum control systems
with complicated Hamiltonians. In this regard, we adopt an
algebraic framework that has been systematically applied to study
atomic structure and molecular spectroscopy
\cite{alhassid1,iachello}. The method is rooted on an intrinsic
symmetry Lie algebra that describes the quantum system under
consideration, whose quantum observables are functions of the
generators of the symmetry algebra. Each of these physical
observables can serve as a control Hamiltonian interacting with
external fields by some realizable physical means, although, not
necessarily under present laboratory conditions. Rather broad
classes of symmetries can be unified in this framework, such as
geometrical, dynamical, or even {\it a priori} prescribed
symmetries \cite{barut}, hence the formulation benefits in gaining
deeper insights into the physical mechanism of quantum control.

In this paper, we are concerned with quantum control systems
associated with finite dimensional symmetry algebras, say
$\la=\{L_1,\cdots,L_d\}_{LA}$. Assume that the system Hamiltonians
$H_0,H_1,\cdots,H_m$ can be expressed as skew-symmetric
polynomials of the generators of $\la$, i.e. elements in the
so-called universal enveloping algebra $E(\la)$ (roughly speaking,
the minimal associative algebra of polynomial operators in terms
of the generators in $\la$ that contains $\la$. see
\cite{onishchik,barut} for rigorous definition). Apparently,
$E(\la)$ is also an infinite dimensional Lie algebra equipped with
the standard definition of Lie bracket $[X,Y]=XY-YX$, where
$X,Y\in E(\la)$. Apparently, the controllability algebra
$\A=\{H_0,H_1,\cdots,H_m\}_{LA}$ is a Lie subalgebra of $E(\la)$.

From the well-known Poincare-Birkhoff-Witt Theorem
(\cite{onishchik}, p.138), all the ordered polynomials
$$\{L_1^{\alpha_1}\cdots
L_d^{\alpha_d};\,\,\alpha_1,\cdots\alpha_d\in\mathbb{N}\}$$
consist of a basis of $E(\la)$. Denote $E^{(n)}(\la)$ the subspace
of elements in $E(\la)$ whose orders are no greater than $n$, we
obtain a graded algebra structure:
$$E^{(1)}(\la){\subset}E^{(2)}(\la){\subset}\cdots{\subset}E^{(n)}(\la)\cdots$$
that decomposes the infinite dimensional vector space $E(\la)$
into finite dimensional subspaces. In this structure, the
computation with differential operators can be replaced by
algebraic operations that are easier to carry out on the graded
finite dimensional subspaces. This facilitates the noncommutative
analysis \cite{nazaikinskii} over an infinite dimensional algebra.

To understand the concepts of symmetry algebra, infinite
dimensional controllability Lie algebra and infinite dimensional
manifold, we present several illustrative examples of quantum
systems with \pt potentials. As one of the known solvable
potentials in the literature \cite{tarn5,alhassid1,iachello}, \pt
potential has been widely used to describe stretching or bending
vibrations states in molecules. In the following, various
algebraic models of \pt potentials will be presented via
separation of variables under special coordinate systems.

{\bf Example 1} The first approach applies the so-called potential
algebra $su(1,1)=\{L_x',L_y',L_z'\}_{LA}$ as the symmetry algebra,
where $L_z'$ is a compact operator and $L_x',L_y'$ are noncompact
operators. Their commutation relations read:
$$[L'_x,L'_y]=iL'_z,\,[L'_y,L'_z]=-iL'_x,\,[L'_z,L'_x]=-iL'_y.$$

\begin{figure}[h]
\centerline{
\includegraphics[width=3.0in,height=2.0in]{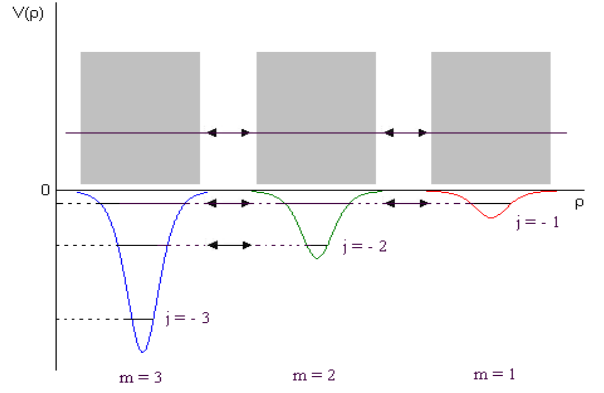}
} \caption{ (Color online) State transitions in example 1 with \pt
potentials.}\label{potential}
\end{figure}

The $su(1,1)$ can be realized in Cartesian coordinates
$$\begin{array}{l}
 L'_x=-i\left(y\frac{\p}{\p{z}}+z\frac{\p}{\p{y}}\right), \\
 L'_y=-i\left(x\frac{\p}{\p{z}}+z\frac{\p}{\p{x}}\right), \\
 L'_z=-i\left(x\frac{\p}{\p{y}}-y\frac{\p}{\p{x}}\right), \\
\end{array}
$$where
$L'_x$ and $L'_y$ are the pseudo angular momentum operators along
$x$ and $y$ axes, and $L'_z$ is the angular momentum operator
along $z$-axis. We change them to the hyperbolic coordinates
$$x=\cosh\rho\cos\phi,\,\,y=\cosh\rho\sin\phi,\,\,z=\sinh\rho,$$
with a succeeding similarity transformation $U=\cosh^{1/2}\rho$ on
the wavefunction. Then, we simultaneously diagonalize the Casimir
operator $C={L'}_x^2+{L'}_y^2-{L'}_z^2$ and ${L'}_z$ with
simultaneous eigenvectors $\{|j,m\rangle=u^m_j(\rho)e^{im\phi}\}$:
$$  C\,|j,m\rangle  =  j(j+1)|j,m\rangle , \quad
  {L'}_z|j,m\rangle  = m|j,m\rangle \\
$$ where the $j$ and $m$ take values in the unitary representations of
$su(1,1)$:
$$
\begin{array}{rl}
  D_j^+: & j=\frac{1}{2},1,\frac{3}{2},\cdots;m=j,j+1,\cdots; \\
  D_j^-: & j=\frac{1}{2},1,\frac{3}{2},\cdots;m=-j,-j-1,\cdots; \\
  C^0_j: & j>0;m=0,\pm{1},\cdots; \\
  C^{1/2}_j: & j>\frac{1}{4};m=\pm\frac{1}{2},\pm\frac{3}{2},\cdots.
\end{array}$$
Finally, we arrive at the time-independent \sd equation subject to
\pt potentials
$$\left(-\frac{d^2}{d\rho^2}-\frac{m^2-\frac{1}{4}}{\cosh^2\rho}\right)u^m_j(\rho)=\left(j+\frac{1}{2}\right)^2u^m_j(\rho).$$
The free Hamiltonian reads $H_0=a(C+\frac{1}{4})$, $a>0$ is some
constant. It possesses discrete (corresponding to $D_j^+$ and
$D_j^-$ representations) or continuous (corresponding to $C_j^0$
and $C_j^{1/2}$ representations) spectra. The potential strength
$m^2-\frac{1}{4}$ is labelled by the eigenvalues of ${L'}_z$.
Choosing control Hamiltonians from the universal enveloping
algebra of the potential algebra, we can realize the quantum
control over this system, for instance, by the other two operators
in the potential algebra \cite{tarn5}:
\begin{equation}\label{pt}
i\hbar{\p\over\p{t}}\psi(t)=\left[\left(C+\frac{1}{4}\right)+u_1{L'}_x+u_2{L'}_y\right]\psi(t).
\end{equation}
Because both the control Hamiltonians commute with the free
Hamiltonian, they are not able to move the energy level labelled
by the eigenvalue $(j+1/2)^2$ of $H_0$. However, as shown in Fig
\ref{potential} \cite{alhassid1}, they can alter the potential
strength because the corresponding operator ${L'}_z$ does not
commute with the control Hamiltonians. Hence these quantum
controls weaken or strengthen potentials while conserving the
system energy. It is not difficult to verify that the
controllability Lie algebra
$\A=\{C+\frac{1}{4},{L'}_x,{L'}_y,{L'}_z\}_{LA}$ is a
four-dimensional Lie subalgebra of $E(su(1,1))$. From the HTC
theorem, this system is strongly analytically controllable.

{\bf Example 2} The second approach describes the scattering
states in \pt potentials. The corresponding symmetry algebra is
called scattering algebra $su(1,1)=\{L_x',L_y',L_z'\}_{LA}$ with
operators $L_y'$ compact and $L_x',L_z'$ noncompact. The
commutation relations read:
$$[L'_x,L'_y]=-iL'_z,\,[L'_y,L'_z]=-iL'_x,\,[L'_z,L'_x]=iL'_y.$$

\begin{figure}[h]
\centerline{
\includegraphics[width=3.0in,height=2.0in]{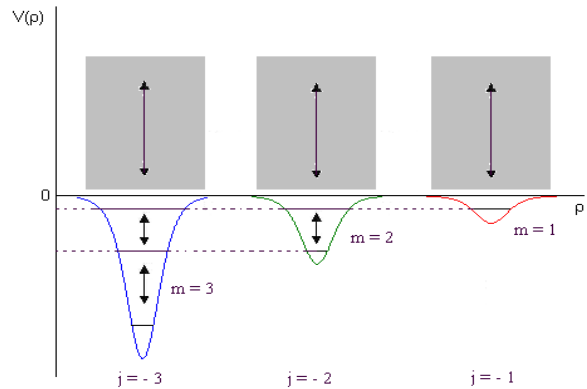}
} \caption{ (Color online) State transitions in examples 2 and 3
under \pt potentials.}\label{energy}
\end{figure}

Here the $su(1,1)$ algebra takes a different realization:
$$\begin{array}{l}
 L'_x=-i\left(y\frac{\p}{\p{z}}+z\frac{\p}{\p{y}}\right), \\
 L'_y=-i\left(x\frac{\p}{\p{z}}-z\frac{\p}{\p{x}}\right), \\
 L'_z=-i\left(x\frac{\p}{\p{y}}+y\frac{\p}{\p{x}}\right), \\
\end{array}
$$where
$L'_x$ and $L'_z$ are the pseudo angular momentum operators along
$x$ and $z$ axes, and $L'_y$ is the angular momentum operator
along $y$-axis. Followed by hyperbolic coordinate transformation
$$x=\cos\varrho\cosh\phi,y=\cos\varrho\sinh\phi,z=\sin\varrho.$$
and a succeeding similarity transformation
$\rho=\tanh^{-1}\cos\varrho$ on the wavefunction, the simultaneous
diagonalization of the operators $C={L'}_x^2-{L'}_y^2+{L'}_z^2$
and ${L'}_z$ with simultaneous eigenvectors
$\{|j,m\rangle=u^m_j(\rho)e^{im\phi}\}$:
$$\begin{array}{rcll}
  C\,|j,m\rangle & = & j(j+1)|j,m\rangle, & j=0,\frac{1}{2},1,\frac{3}{2},2,\cdots \\
  {L'}_z|j,m\rangle & = & m|j,m\rangle, & m\in\mathbb{R},\\
\end{array}$$
gives the time-independent \sd equation subject to \pt potential
$$\left(-\frac{d^2}{d\rho^2}-\frac{j(j+1)}{\cosh^2\rho}\right)u^m_j(\rho)=m^2u^m_j(\rho).$$
The free Hamiltonian reads $H_0=a{L'}_z^2$, where $a>0$ is a
constant. The system possesses a positive continuous spectrum
$\{E=am^2,\,m\in\mathbb{R}\}$ because ${L'}_z$ is non-compact. The
potential strength $j(j+1)$ is related to the Casimir operator
$C$. Consider quantum control system using two operators in the
scattering algebra \cite{tarn5}:
\begin{equation}\label{st}
i\hbar{\p\over\p{t}}\psi(t)=[a{L'}_z^2+u_1{L'}_x+u_2{L'}_y]\psi(t),
\end{equation}
the potential strength is a conservative quantity because the
corresponding Casimir operator commutes with all system
Hamiltonians. As shown in the upper part of Fig \ref{energy}
\cite{alhassid1}, the controls affect only the change of energies
in continuous spectra. The controllability algebra
$\A=\{a{L'}_z^2,{L'}_x,{L'}_y\}_{LA}$ is an infinite dimensional
Lie algebra and contains arbitrarily high-order elements in
$E(su(1,1))$. Therefore, the former results cannot be used here.

{\bf Example 3} The third example describes bound states in \pt
potentials with a compact dynamical Lie algebra
$su(2)=\{L_x',L_y',L_z'\}_{LA}$. In a similar way, we firstly
realize this algebra in Cartesian coordinate system:
$$\begin{array}{l}
 L'_x=-i\left(y\frac{\p}{\p{z}}-z\frac{\p}{\p{y}}\right), \\
 L'_y=-i\left(x\frac{\p}{\p{z}}-z\frac{\p}{\p{x}}\right), \\
 L'_z=-i\left(x\frac{\p}{\p{y}}-y\frac{\p}{\p{x}}\right), \\
\end{array}
$$
where $L'_j$ is the angular momentum operator along $j$-axis,
$j=x,y,z$. Changing them into spherical coordinates
$$x=\cos\varrho\cos\phi,y=\cos\varrho\sin\phi,z=\sin\varrho$$followed
by a similarity transformation $\rho=\cos^{-1}\sech\varrho$ on the
wavefunction, and the simultaneous diagonalization of the
operators $C={L'}_x^2+{L'}_y^2+{L'}_z^2$ and ${L'}_z$ with
simultaneous eigenvectors $\{|j,m\rangle=u^m_j(\rho)e^{im\phi}\}$
$$\begin{array}{rcll}
  C|j,m\rangle & = & j(j+1)|j,m\rangle, & j=\frac{1}{2},1,\frac{3}{2},2,\cdots \\
  {L'}_z|j,m\rangle & = & m|j,m\rangle, & m=-j,\cdots,j,\\
\end{array}$$
we can get the time-independent \sd equation subject to \pt
potentials
$$\left(-\frac{d^2}{d\rho^2}-\frac{j(j+1)}{\cosh^2\rho}\right)u^m_j(\rho)=-m^2u^m_j(\rho).$$
The free Hamiltonian reads $H_0=-a{L'}_z^2$, where $a>0$ is a
constant. The potential strength is labelled by the eigenvalues of
the Casimir operator, while the quantum number $m$ labels the
$(2j+1)$ bound states in the $j$-th potential. Consider the
following control system \cite{dong2}:
\begin{equation}\label{bt}
i\hbar{\p\over\p{t}}\psi(t)=[a{L'}_z^2+u_1L'_x+u_2L'_y]\psi(t),
\end{equation}
with control Hamiltonians selected from the $su(2)$ algebra. As
shown in the lower part of Fig \ref{energy}, the control
Hamiltonians serve to shift discrete energy levels in \pt
potentials. They cannot affect the potential strength, nor can
they drive the system out of discrete spectra to the continuum.
More interestingly, this example shows that controllability
algebra can still be infinite dimensional even if the symmetry
algebra is compact.

The above three examples reveal rich symmetry properties in the
same \pt potentials by which system can be represented by simple
algebraic variables. The concepts of infinite dimensional Lie
algebra and infinite dimensional manifold should not be confused
with the infinite dimensional systems (distributed-parameter
systems \cite{ball}) widely used in the literature. All the three
examples are infinite dimensional systems since they are described
by partial differential equations and their solutions are
functions. However, in the first example, the controllability
algebra is finite dimensional and the corresponding manifold is a
finite dimensional submanifold of infinite dimensional manifolds.
This is different from the finite dimensional manifolds contained
in a finite dimensional vector space, since the generators of the
algebras consists of unbounded operators and the solution of the
\sd equations are functions rather than vectors in finite
dimensional vector spaces.

\section{Smooth Controllability}

No doubt that the domain problem is also non-trivial for systems
with infinite dimensional controllability Lie algebras. Normally,
it is not difficult to find an analytic domain for the symmetry
algebra $\la$, however, this domain is not invariant under the
actions of high-order operators in its universal enveloping
algebra $E(\la)$. In fact, it has been demonstrated that analytic
domain is nonexistent for most infinite dimensional Lie algebras
\cite{arnal}. Therefore, one needs to seek another proper domain
for the universal enveloping algebra.

What we are going to choose is the larger set of differentiable
vectors of $\la$:
\begin{equation}\label{ds}
\Ds=\{\phi\in\h : \|L_1^{s_1}\cdots L_d^{s_n}\phi\|<\infty;
\forall s_i=0,1,2,\cdots\},
\end{equation}
which is well-defined and invariant for all operators in $E(\la)$,
hence can be taken as a candidate. Parallel with the analytic
domain, we call $\Ds$ the smooth domain. Topologically, the smooth
domain occupies a special ILH (inverse limit Hilbert) vector space
structure that is inherited from, but completely different with,
the Hilbert space structure of $\h$ (see Appendix \ref{ilh}). This
ILH structure also makes it possible to generate an infinite
dimensional ILH transformation group on $\h$ from the universal
enveloping algebra $E(\la)$ \cite{omori}, which establishes a
solid mathematical basis for the following discussions on infinite
dimensional quantum control systems.

Moreover, the smooth domain has crucial physical meanings on
describing scattering states, which can be intuitively depicted as
limits of bound states whose wavepackets spread widely in the
configuration space. By definition, $\Ds$ represents the set of
bound states on which all the system observables (elements in the
universal enveloping algebra) and their commutations act legally,
hence is comprised of all experimentally preparable states
produced by the quantum control system (\ref{abc}). Denote the
dual space of $\Ds$ by $\Ds^*$, i.e. the space of continuous
antilinear functionals on $\Ds$. The scattering states can be
represented as ideal vectors in the larger set $\Ds^*$ that do not
have finite norm in $\h$ \cite{melsheimer,madrid}. Mathematically,
the triad of linear spaces
$$\Ds\subset\h\subset\Ds^*$$
forms a so-called rigged Hilbert space \cite{melsheimer,madrid}
with $\Ds$ dense in $\h$ and $\h$ dense in $\Ds^*$ (with respect
to the weak topology of $\Ds^*$). So $\Ds$ is also dense in
$\Ds^*$. This shows that scattering states can be identified as
the ideal limits of some sequences of bound states in $\Ds$ ({\it
N.B.} the limit is called "ideal" because it is outside the
Hilbert space and is not converged to with respect to the usual
Hilbert space topology).

Consider the manifold $\M$ defined by (\ref{m}), which,
corresponding to infinite dimensional controllability algebras, is
in general infinite dimensional. Parallel with the analytic
controllability, we present rigorous controllability concepts of
quantum control systems over this manifold:
\begin{definition}{\bf (Smooth Controllability)}
Quantum mechanical control system (\ref{abc}) is said to be
smoothly controllable if the reachable set $\R(\psi)=\M\cap\Ds$
for all $\psi\in\M\cap\Ds$.
\end{definition}
\begin{definition}{\bf (Strong Smooth Controllability)}
The system is said to be strongly smoothly controllable if the
reachable set $\R_t(\psi)=\M\cap\Ds$ for all $\psi\in\M\cap\Ds$ at
any time $t>0$.
\end{definition}

In the context of this paper, where only piecewise constant
controls are considered, the system can never be completely
smoothly controllable on an infinite dimensional manifold $\M$
unless an infinite number of switches are applied \cite{tarn2}.
Hence we have to turn to the following weakened definition:
\begin{definition}{\bf (Approximate Smooth Controllability)} Quantum
mechanical control system (\ref{abc}) is said to be approximately
smoothly controllable on $\M$ if the reachable set $\R(\psi)$ is
dense (with respect to the ILH-topology) in $\M\cap\Ds$ for all
$\psi\in\M\cap\Ds$.
\end{definition}
\begin{definition}\label{smc}{\bf (Approximate Strong Smooth
Controllability)} The system is said to be approximately strongly
smoothly controllable if the reachable set $\R_t(\psi)$ is dense
in $\M\cap\Ds$ (with respect to the ILH-topology) for all
$\psi\in\M\cap\Ds$ at any time $t>0$.
\end{definition}

The consideration of differential vectors has been suggested by
Huang, Tarn and Clark in \cite{tarn2} as a possible extension of
the analytic controllability. In this paper, we put it into a
rigorous setting. The smooth domain not only enlarges the system
domain, but also gives a nice topology by which one can carry out
strict controllability analysis. More important a byproduct is
that the smooth domain provides an explicit picture of the control
of scattering-state quantum systems, which are {\it non-physical}
in a strict sense. Nevertheless, taking scattering states as ideal
limits of bound states in the smooth domain, we can directly
translate the knowledge of controllability properties on the
smooth domain $\Ds$ to scattering states in the dual space $\Ds^*$
in the sense of ideal limit.

\section{Main Results}

The foregoing mathematical preliminaries provide a technical basis
for us to explore quantum systems with infinite dimensional
controllability algebras. In this section, we will study the
approximate strong smooth controllability of quantum control
systems over the manifold $\M$ defined by (\ref{m}). The main
difficulty will be that only a finite number of controls are
available to guide the quantum states all over the infinite
dimensional manifold $\M$, while fighting against the possibly
complex free evolution.

In the following, the discussion will be focused on the quantum
control system of the unitary propagators:
\begin{equation}\label{qcu}
\frac{\dd }{\dd
t}U(t)=\left[H_0+\sum_{j=1}^mu_j(t)H_j\right]U(t),\,\,\,U(0)=I,
\end{equation}
where the system propagator satisfies $\psi(t)=U(t)\psi_0$ and
belongs to the infinite dimensional Lie group $\G$ generated by
the controllability algebra $\A$. Controllability properties of
system (\ref{abc}) can be derived from this system, because the
manifold $\M$ can be equivalently expressed as the orbit,
$\M=\{U\psi_0,\,U\in\G\}$, of $\G$ passing the initial state
$\psi_0$. Providing that a dense subset of propagators in $\G$ can
be generated from (\ref{qcu}), the system will be consequently
smoothly controllable on $\M$. Our basic idea is to construct
unitary transformations in $\G$ by repeatedly switching control
interactions so that the free evolution can be cancelled and
recreated, hence forms a dense subset of $\G$ as the number of
switches increase. Concretely, we are looking for a class of
strongly "adjustable" flows (i.e. one-dimensional subgroups in
$\G$) generated by the so-called strongly attainable Hamiltonians:
\begin{definition} Denote the infinitesimal-time reachable set
$\R_0(\psi)=\bigcap_{t>0}\R_{\leq t}(\psi)$, where $\R_{\leq
t}(\psi)$ denotes the reachable set within $t$ units of time. A
time-independent Hamiltonian $X$ is said to be strongly attainable
if its integral curve $\{\exp(Xt)\psi,\,t\in\mathbb{R}\}$ passing
any $\psi\in\M\cap\Ds$ is contained in the closure of
$\R_0(\psi)$. A set of Hamiltonians is said to be strongly
attainable if each element in this set is strongly attainable.
\end{definition}

The strongly attainable Hamiltonians represent Hamiltonians of
which the generated unitary propagators can be achieved in an
arbitrary small time interval. We are going to seek plenty of
strongly attainable Hamiltonians for the system (\ref{qcu}) so
that flows passing $\psi_0$ can be generated to steer the system
with full freedom in a dense subset of the infinite dimensional
manifold $\M$, which leads to the approximate strong smooth
controllability of (\ref{abc}). Before presenting the main result,
we will prove several preliminary lemmas and theorems. Readers
interested in the main result may proceed directly to Theorem
\ref{ac} and examples in section \ref{example}.

To apply these ideas, we will repeatedly use the
Campbell-Baker-Hausdorff formula that is already well established
on infinite dimensional Lie groups \cite{onishchik,teichmann}:
\begin{eqnarray*}
  X_s\cdot Y_t &=& \Exp\left(sX+tY+\frac{st}{2}[X,Y]+\frac{s^2t}{12}[X,[X,Y]]\right. \\
   && \left. +\frac{st^2}{12}[Y,[Y,X]]+\cdots\right)
\end{eqnarray*}
and the Trotter's formula \cite{trotter,davis}:
\begin{eqnarray}
  (X+Y)_s\phi &=&
   \label{trotter1}  \lim_{n\rightarrow\infty}(X_{\frac{s}{n}}Y_{\frac{s}{n}})^n\phi,\\
\label{trotter2}   [X,Y]_s\phi &=&
\lim_{n\rightarrow\infty}(X_{\sqrt{\frac{s}{n}}}Y_{\sqrt{\frac{s}{n}}}X^{-1}_{\sqrt{\frac{s}{n}}}Y^{-1}_{\sqrt{\frac{s}{n}}})^n\phi,
\end{eqnarray}
where $X_s$ and $Y_t$ are the flows generated by $X$ and $Y$
respectively. Firstly, one can identify the following properties
of strongly attainable Hamiltonians:

\begin{lemma}\label{l1}
Let $X$ and $X_1,\cdots,X_\alpha$ be strongly attainable
Hamiltonians, then
\begin{enumerate}
    \item $(X_{i_n})_{s_n}\cdots(X_{i_1})_{s_1}\psi\in\overline{\R_0(\psi)}$, for all $n\in\mathbb{N}$, $i_1,\cdots,i_n\in\{1,\cdots,\alpha\}$, $s_1,\cdots s_n\in\mathbb{R}$
    and $\psi\in\Ds$;
    \item $(H_0)_tX_s\psi\in\overline{\R_t(\phi)}$, $X_s(H_0)_t\psi\in\overline{\R_t(\phi)}$, for all
    $\psi\in\R_t(\phi)$, $t>0$, and $s\in\mathbb{R}$.
\end{enumerate}
\end{lemma}
\noindent {\bf Proof}: Let $U(t,{\bf u}(\cdot))$ be the system
propagator under control ${\bf
u}(\cdot)=(u_1(\cdot),\cdots,u_m(\cdot))$. According to the
definition of strongly attainable Hamiltonians, for arbitrary
positive number $\epsilon$ and arbitrary positive time $t$, there
exist a sequence of unitary propagators
$U_{1,t/n},\cdots,U_{n,t/n}$ such that $U(t/n,{\bf
u}^k(\cdot))=U_{k,t/n}$ for some ${\bf u}^k(\cdot)$, and satisfies
$$\left\|[(X_{i_k})_{s_k}-U_{k,t/n}](X_{i_{k-1}})_{s_{k-1}}\cdots
(X_{i_1})_{s_1}\psi\right\|<\frac{\epsilon}{n}.$$ Because
$U_{k,t/n}$'s are unitary operators,
\begin{widetext}
\begin{eqnarray*}
   && \|(X_{i_n})_{s_n}(X_{i_{n-1}})_{s_{n-1}}\cdots
(X_{i_1})_{s_1}\psi-U_{n,t/n}U_{n-1,t/n}\cdots U_{1,t/n}\psi\| \\
  &<&
  \|[(X_{i_n})_{s_n}-U_{n,t/n}](X_{i_{n-1}})_{s_{n-1}}\cdots
(X_{i_1})_{s_1}\psi\|+\cdots+\|U_{n,t/n}U_{n-1,t/n}\cdots[
(X_{i_1})_{s_1}- U_{1,t/n}]\psi\|\\
  &=&
  \|[(X_{i_n})_{s_n}-U_{n,t/n}](X_{i_{n-1}})_{s_{n-1}}\cdots
(X_{i_1})_{s_1}\psi\|
  +\cdots+\|(X_{i_1})_{s_1}\psi-U_{1,t/n}\psi\|\\
  &<&\epsilon
\end{eqnarray*}\end{widetext}

From the choices of $U_{1,t/n},\cdots,U_{n,t/n}$, it is obvious
that $U_{n,t/n}U_{n-1,t/n}\cdots U_{1,t/n}\psi\in\R_t(\psi)$ for
any $t>0$. So $(X_{i_k})_{s_k}(X_{i_{k-1}})_{s_{k-1}}\cdots
(X_{i_1})_{s_1}\psi\in\overline{\R_t(\psi)}$. As to the second
assertion, we similarly choose $U_\tau$ such that
$\|(H_0)_tX_s\psi-(H_0)_tU_\tau\psi\|<\epsilon$. Then
$$\begin{array}{ll}
  &\|(H_0)_tX_s\psi-(H_0)_{t-\tau}U_\tau\psi\| \\
<& \|(H_0)_tX_s\psi-(H_0)_tU_\tau\psi\| +\|(H_0)_tU_\tau\psi-(H_0)_{t-\tau}U_\tau\psi\| \\
\end{array}\\
$$
Because of the continuity of the one-parameter group
$\{(H_0)_t\}_{t\in\mathbb{R}}$, the second term goes to zero as
$\tau\rightarrow 0$. Therefore
$(H_0)_tX_s\psi\in\overline{\R_t(\psi)}$. Similar is the proof for
$X_s(H_0)_t\psi\in\overline{\R_t(\psi)}$

\begin{lemma}\label{l2}
$X$ is strongly attainable if and only if
$(H_0+cX)_t\psi\in\overline{\R_t(\psi)}$ for all $c\in\mathbb{R}$.
\end{lemma}

\noindent {\bf Proof}: To prove the sufficiency, we estimate the
deviation between the system evolution with that in absence of
$H_0$ by integrating the differential equation
$$\frac{d}{ds}(\e{H_0}+X)_{t-s}X_s\psi=-(\e{H_0}+X)_{t-s}\left(\e{H_0}\right)X_s\psi$$
over the time interval $[0,t]$:
\begin{equation}
\begin{array}{cl}
   & \|(\e{H_0}+X)_{t}\psi-X_t\psi\| \\
  \leq & \int_0^t{\|(\e{H_0}+X)_{t-s}\|\cdot\|\left(\e{H_0}\right)X_s\psi\|{\dd}s} \\
  \leq & \e{Mt}
\end{array}
\end{equation}
where $\|(\e{H_0}+X)_{t-s}\|=1$ by unitarity,
$0<M=\sup_{[0,t]}\|H_0X_s\psi\|<\infty$. Hence for fixed $t$ and
$\psi$,
$\lim_{\e\rightarrow{0}}\|(\e{H_0}+X)_{t}\psi-X_t\psi\|=0$. This
is to say, the integral curve $X_t\psi$ can be arbitrarily close
to
$(\e{H_0}+X)_{t}\psi=(H_0+\e^{-1}X)_{\e{t}}\psi\in\overline{\R_{\e{t}}(\psi)}$
for $\e$ small enough. Therefore
$X_t\psi\in\overline{\R_0(\psi)}$, i.e. $X$ is strongly
attainable.

Conversely, if $X$ is strongly attainable, the necessity can be
shown from the Trotter's formula
$$(H_0+cX)_s\psi=\lim_{n\rightarrow\infty}[(H_0)_{s/n}(cX)_{s/n}]^n\psi,$$
in which the term at the right hand side belongs to the closure of
$\R_t(\psi)$ by repeatedly using Lemma \ref{l1} (2). The end of
proof.

\begin{lemma}\label{l3}
Denote $\A_S$ the collection of strongly attainable Hamiltonians.
$\A_S$ is a Lie algebra containing $\B=\{H_1,\cdots,H_r\}_{LA}$.
\end{lemma}
\noindent{\bf Proof}: Suppose $X,Y\in\A_S$ is strongly attainable.
The strong attainability of $cX$ for arbitrary nonzero
$c\in\mathbb{R}$ is obvious. Using the Trotter's formula and
applying Lemma \ref{l1} (1), we observe that the right hand sides
of Eqs.(\ref{trotter1}) and (\ref{trotter2}) belongs to the
closure of the infinitesimal-time reachable set, so both $X+Y$ and
$[X,Y]$ are strongly attainable. This proves the Lie algebra
property of $\A_S$. On the other hand, the control Hamiltonians
$H_1,\cdots, H_m$ are strongly attainable according to Lemma
\ref{l2}. Therefore the Lie algebra $\B$ that is generated by
$H_1,\cdots,H_m$ is a strongly attainable Lie subalgebra of
$\A_S$.

\begin{theorem}\label{c}
The Lie algebra $\C$ is strongly
attainable if the algebraic condition $[\B,\,\C]\subseteq\B$ is
satisfied.
\end{theorem}

\noindent {\bf Proof}: It is sufficient to prove that
$\ad_{H_0}^kH\in\A_S$ for arbitrary integer $k$ and strongly
attainable Hamiltonian $H\in\B$. We invoke the
Campbell-Baker-Hausdorff formula \cite{onishchik}:
\begin{widetext}
$$ H_{-t}(H_0)_sH_t\psi
  =
\left(H_0+t\,\ad_{H_0}H+\frac{t^2}{2}\int_0^1{(\theta-1)^{2}H_{-\theta
t}\ad_H^2H_0H_{\theta t}{\dd}\theta}\right)_s\psi,
$$\end{widetext} where $t\in\mathbb{R}$. The last term
$$R_1=\frac{t^2}{2}\int_0^1{(\theta-1)^{2}H_{-\theta
t}\ad_H^2H_0H_{\theta t}{\dd}\theta}$$ is the Lagrange remainder.
Under the condition $[\,\B,\C\,]\subseteq{\B}$, the term
$\ad_H^2H_0$ is strongly attainable, hence its translation
$H_{-\theta t}\ad_H^2H_0H_{\theta t}$ by a strongly attainable
Hamiltonian $H$ is also strongly attainable (by lemma \ref{l1}).
Taking the integral as a limit of summations, we can see that the
Lagrange remainder is also strongly attainable. Applying Trotter's
formula, we have
$$(H_0+t{\ad}_{H_0}H)_s\psi=\lim_{n\rightarrow \infty}[H_{-t}(H_0)_{s/n}H_t(-R_1)_{s/n}]^n\psi,$$
of which the right hand side is contained in the closure of
reachable set $\R_s(\psi)$ by Lemma \ref{l1}. Hence
$(H_0+t{\ad}_{H_0}H)_s\psi\in\overline{\R_s(\psi)}$, which implies
that $\ad_{H_0}H$ is strongly attainable from Lemma \ref{l2}. Thus
we proved the strong attainability of ${\ad}_{H_0}\B$.

Inductively, assume the subspace of Hamiltonians $\C_k=\{{\rm
ad}_{H_0}^j\B,j=0,\cdots,k\}\subset \C$ is strongly attainable for
some positive integer $k$. Employ again the
Campbell-Baker-Hausdorff formula:
\begin{widetext}
\begin{eqnarray}
\nonumber &&(\ad_{H_0}^kH)_{-t}(H_0)_s(\ad_{H_0}^kH)_t\psi \\
\nonumber &=&
\left(H_0-t\,\ad_{\ad_{H_0}^kH}H_0+\frac{t^2}{2}\int_0^1{(\theta-1)^{2}(\ad_{H_0}^kH)_{-\theta
t}\ad_{\ad_{H_0}^kH}^2H_0(\ad_{H_0}^kH)_{\theta
t}{\dd}\theta}\right)_s\psi,\\
 &=&
\left(H_0+t\,\ad_{H_0}^{k+1}H+\frac{t^2}{2}\int_0^1{(\theta-1)^{2}(\ad_{H_0}^kH)_{-\theta
t}\left[\ad_{H_0}^{k+1}H,\ad_{H_0}^kH\right](\ad_{H_0}^kH)_{\theta
t}{\dd}\theta}\right)_s\psi. \label{cbhk}
\end{eqnarray}
\end{widetext}
Making use of the formula \cite{kunita}
\begin{eqnarray*}
   && \left[\ad_{H_0}^{k+1}H,\ad_{H_0}^kH\right] \\
   &=& \sum_{j=0}^k{(-1)^j\left(%
\begin{array}{c}
  k\\
  j \\
\end{array}%
\right)\ad_{H_0}^{k-j}\left[\ad_{H_0}^{k+j+1}H,H\right]}\subseteq{\C_k}
\end{eqnarray*}
together with the condition $[\,\B,\C\,]\subseteq{\B}$, we can see
the strong attainability of the Lagrange remainder $R_k$ in
(\ref{cbhk}) from the assumption that $\C_k$ is strongly
attainable. Using Trotter's formula, we have
\begin{eqnarray*}
   && (H_0+t{\ad}_{H_0}^{k+1}H)_s\psi \\
   &=& \lim_{n\rightarrow \infty}\left[(\ad_{H_0}^kH)_{-t}(H_0)_{s/n}(\ad_{H_0}^kH)_{t}(-R_k)_{s/n}\right]^n\psi,
\end{eqnarray*}
of which the right hand side is contained in the closure of
reachable set $\R_s(\psi)$ by Lemma \ref{l1}. Similarly, we can
prove in the same line as carried above that each element in the
subspace ${\rm ad}_{H_0}^{k+1}\B$ is strongly attainable.
Therefore, we inductively prove the strong attainability of
$\C_{k+1}$. In conclusion, $\C$ is strongly attainable.

Next, we cite a useful theorem to connect the strongly attainable
Hamiltonians with the reachable sets $\R_t(\psi)$:

\begin{theorem} \cite{sj,kunita,hirschorn1}\label{s}
Let $I(\psi,\C)=(\Exp\C)\psi$ be the maximal connected integral
manifold of $\C$ containing the point $\psi$. Then $\R_t( \psi )
\subseteq I^t(\psi,\C)$, where
$I^t(\psi,\C)=(H_0)_tI(\psi,\C)=I((H_0)_t\psi,\C)$.
\end{theorem}
\noindent {\bf Proof}: Because the admissible control are
piecewise-constant, we can always decompose the system flow into
pulses driven by constant controls. Consider the single pulse, we
have by Trotter's formula and Campbell-Baker-Hausdorff formula:
\begin{eqnarray}
\nonumber   && (H_0+H)_t\psi   \\
\nonumber   &=&
   \lim_{n\rightarrow\infty}[(H_0)_{\frac{t}{n}}H_{\frac{t}{n}}]^n\psi    \\
   &=& \lim_{n\rightarrow\infty}(H_0)_t
\left\{(H_0)_{-\frac{n-1}{n}t}H_{\frac{t}{n}}(H_0)_{-\frac{n-1}{n}t}\right\}\\
&&\cdots\left\{(H_0)_{-\frac{t}{n}}H_{\frac{t}{n}}(H_0)_{\frac{t}{n}}\right\}(H_0)_{-\frac{t}{n}}\psi
\label{curly}
\end{eqnarray}
where $H=\sum_{i=1}^m{u_iH_i}$ are the control Hamiltonian.
Because
\begin{eqnarray*}
   && (H_0)_{-s}H_t(H_0)_s\psi \\
   &=& \Exp\left\{tH-st[H_0,H]+\frac{s^2t}{2}[H_0,[H_0,H]]-\cdots\right\}\psi\\
   &\in&\Exp \C\psi,
\end{eqnarray*} so each part in the curly brackets in (\ref{curly})
generates a unitary propagator in $\Exp\C$. Hence $\R_t( \psi )
\subseteq I^t(\psi,\C)$. Inductively, suppose the conclusion
establishes for $k$ pulses, then for $(k+1)$ pulses with
$t_1+\cdots +t_{k+1}=t$, let $t'=t_1+\cdots +t_{k}$,
\begin{eqnarray*}
   && (H_0+H^{(k+1)})_{t_{k+1}}(H_0+H^{(k)})_{t_{k}}\cdots(H_0+H^{(1)})_{t_{1}}\psi \\
  &\in& [\Exp\C(H_0)_{t_{k+1}}] (H_0)_{t'}I(\psi,\C)\\
  &\in& \Exp\C (H_0)_{t} I(\psi,\C)\\
  &=&(H_0)_{t}[(H_0)_{-t}\Exp\C(H_0)_{t}]I(\psi,\C)=(H_0)_{t}I(\psi,\C)
\end{eqnarray*}
Thus the conclusion establishes for $(k+1)$ pulses. The end of
proof.

Having the above properties of the strongly attainable
Hamiltonians and reachable sets in hand, now we can draw the main
result of the approximate strong smooth controllability:
\begin{theorem}\label{ac}
The system (\ref{abc}) is approximately strongly smoothly
controllable if the following conditions are satisfied:
\begin{enumerate}
    \item $[\,\B,\C\,]\subseteq{\B}$;
    \item For any $\phi\in\M\cap\Ds$, $\C(\phi)=\A(\phi)$ and they
    are infinite dimensional.
\end{enumerate}
\end{theorem}

\noindent{\bf Proof}: From Theorem \ref{s}, $\C$ is strongly
attainable under the first condition. According to Lemma \ref{l1},
we have $I^t(\psi,\C)=(H_0)_t\Exp \C \psi \subseteq
\overline{\R_t(\psi)}$ for some initial state $\psi\in\M\cap\Ds$.
On the other hand, from Theorem \ref{s},
$\overline{\R_t(\psi)}\subseteq I^t(\psi,\C)$. Hence
$\overline{\R_t(\psi)}=I^t(\psi,\C)$.

By Frobenius theorem (\cite{omori}, p.215), the condition
$\C(\phi)=\A(\phi)$ for any $\phi\in\M\cap\Ds$ guarantees that
$\A$ and $\C$ have identical maximal integral manifold passing
$\psi$, i.e. $\Exp \A \psi=\Exp \C\psi=\M\cap\Ds$. Since
$H_0\in\A$, the unitary transformation $(H_0)_t$ leaves $\M$
invariant.Thus we arrive at the final conclusion:
$$\overline{\R_t(\psi)}=(H_0)_t(\Exp \C)\psi=(H_0)_t(\M\cap\Ds)=\M\cap\Ds.$$

\section{Examples}\label{example}

To illuminate the ideas presented in this paper, we proceed to
discuss several examples in this section.

{\bf Example 1} The first paradigm comes from the model of
continuous quantum computation over continuous variables proposed
by Lloyd and Braunstein \cite{lloyd4}. The scheme encodes quantum
information in the continuous spectrum of the position operator of
a harmonic oscillator. The control task then becomes the
manipulation of superpositions of eigenstates of the position
operator by the following control systems:
\begin{equation}\label{lloyd}
\begin{array}{cl}
  i{\p\over\p{t}}\psi(x,t) = & [p^2+x^2+u_1(xp+px)+u_2p+u_3x \\
   & +u_4(x^2+p^2)^2]\psi(t) \\
\end{array}, \\
\end{equation}
where the commutation of the position operator $x$ and the
momentum operator $p=-i\hbar\p_x$ reads $[x,p]=i\hbar$. Here the
Heisenberg algebra $h(1)=\{x,p,i\}_{LA}$ plays the role of the
symmetry algebra of the system. The smooth domain for the
Heisenberg algebra $h(1)=\{x,p,i\}$ is the Schwartz space
$$\left\{v(x)\in{L^2(\mathbb{R})}:\sup_{\alpha,\beta\geq{0}}{\biggl |}x^\alpha\left(\frac{d}{dx}\right)^\beta{v(x)}{\biggl |}<\infty\right\}.$$
As argued in \cite{lloyd4}, arbitrary functions of variable $x$
and $p$ can be approximated by repeatedly switching operations of
the control Hamiltonians, i.e. all the polynomials of $x$ and $p$
in $E(h(1))$ can be generated by commutations and linear
combinations of the control operators
$$H_1=xp+px,\,\,H_2=p,\,\,H_3=x,\,\,H_4=(x^2+p^2)^2.$$
This is actually equivalent to say that all these polynomial
Hamiltonians are strongly attainable according to the terminology
used in this paper. According to Theorem \ref{ac}, the system
(\ref{lloyd}) is approximately strongly smoothly controllable. In
quantum computation field, this amounts to the universality of
continuous quantum computation using model (\ref{lloyd}).

Physically, the linear optical interactions are used to shift
phase by $x$ and translate the coordinate by $p$; the second-order
operator $px+xp$ provides a squeezer operation. The nonlinear Kerr
Hamiltonian $(x^2+p^2)^2$, which plays essential roles in
producing many interesting physical phenomena such as entangled
photons, is applied here to explode up an infinite dimensional
controllability algebra that is necessary for controllability over
the whole continuous spectrum. It is also easy to verify that many
other higher-order operators in $E(h(1))$ can replace the Kerr
Hamiltonian for the same goal of controllability \cite{lloyd4}.

{\bf Example 2} The physical model of the second example has been
described in section \ref{problem}. The scattering states in \pt
potentials are characterized using a noncompact symmetry algebra
$su(1,1)$. Let $|j,k\rangle$, $k=j,j+1,\cdots$, be simultaneous
eigenvectors of the compact generator ${L'}_y$ and the Casimir
operator $C$ for some fixed integer $j>0$. These vectors expand a
Hilbert space $\h_j$. The smooth domain contained in $\h_j$
consists of the "fast decreasing sequences"
\begin{equation}
\Ds=
\{x=\sum_{k=j}^\infty{x_k|j,k\rangle}\,|\,\lim_{|k|\rightarrow\infty}k^nx_k=0,\,\forall\,n\in\mathbb{N}\}
\end{equation}
as described in \cite{lindblad}. The scattering states are
contained in the set of "slow increasing sequences"
\begin{equation}
\Ds^*=
\{x=\sum_{k=j}^\infty{x_k|j,k\rangle}\,|\,\lim_{|k|\rightarrow\infty}k^{-n}x_k=0,\,\forall\,n\in\mathbb{N}\}
\end{equation}

For the quantum control system (\ref{st}), one can verify
inductively that the controllability algebra $\A=E(su(1,1))$ (see
proof in Appendix \ref{su11}), which generates a unitary
representation of the volume-preserving diffeomorphism group
$\Df{S^{1,1}}$ over a hyperboloid surface
$S^{1,1}=SU(1,1)/SO(1,1)$ \cite{chern}. Denote $\M$ the orbit of
$\Df{S^{1,1}}$ passing the initial state $\psi\in\Ds\cap\h_j$.

However, although we have an infinite dimensional controllability
algebra $\A$, the algebra $\B=\{{L'}_x,{L'}_y\}_{LA}=su(1,1)$ is
too small to fulfil the condition in Theorem \ref{ac}. Hence
nothing can be told according to the results obtained in this
paper. Nevertheless, if one can apply an extra second-order
control Hamiltonian ${L'}_x^2$, which leads to the following
control system:
\begin{equation}\label{st1}
i\hbar{\p\over\p{t}}\psi(t)=[a{L'}_z^2+u_1L'_x+u_2L'_y+u_3{L'}_x^2]\psi(t),
\end{equation}
the second-order control Hamiltonian helps to explode up an
infinite dimensional Lie algebra $\B$ that coincides with
$\A=E(su(1,1))$. According to Theorem \ref{ac}, strong approximate
smooth controllability follows on $\M$.

Let us give some physical insights. At the first glance, the
operators ${K'}_\pm={L'}_x\pm{{L'}_y}$ resemble the ladder
operators in harmonics oscillators. But intuitively, the "ladder"
operators are not allowed to generate discrete shift of levels on
a continuous spectrum. In fact, ${K'}_\pm$ shift eigenvalues of
${L'}_z$ by $\pm i$ units \cite{mukunda}, which is of course
absurd because the Hermitian ${L'}_z$ has a real continuous
spectrum. As interpreted in \cite{mukunda,mukunda2}, the
contradiction originates from the fact that ${K'}_\pm$ act
illegally upon the scattering states that are outside the Hilbert
space. Their operations are only well-defined on the wavepackets
of superposition of scattering states. Interested readers may
refer to \cite{mukunda,mukunda2} for more details.

The second-order Hamiltonian ${L'}_x^2=({K'}_++{K'}_-)^2/4$ is
also essential in expanding an infinite dimensional
controllability algebra as well as the Kerr Hamiltonian in the
first example. Any other second-order operator that does not
commute with $H_0$ functions equivalently in controllability. As
analogs of the Kerr nonlinear process, these operators generate
higher-order harmonics on $SU(1,1)$. On the other hand, whatever
higher-order operators in $E(su(1,1))$ are applied, they can never
move the system state out of the continuous spectrum due to the
symmetry pre-determined by the scattering algebra.

\section{Conclusion}
This paper provides a clearer understanding of system control of
infinite dimensional quantum mechanical systems. The presented
framework may be applied to quantum control systems with finite or
infinite dimensions, and with bound states or scattering states.
Back to the cases of finite dimensional controllability algebras,
the extension of analytic controllability to the larger smooth
domain, which has been conjectured in \cite{tarn2}, can be taken
as a corollary of Theorem \ref{ac}. Most important is that the
results open up much broader applications to infinite dimensional
manifolds.

As has been earlier discussed by Zhao and Rice \cite{rice3},
control of scattering-state system can be significantly influenced
by the presence of chaos. Since the strong controllability
property is not altered because the evolution always concerns
itself on finite time intervals, our results affirm that quantum
scattering-state control system can be "strong" enough to overcome
the chaos. However, controllability not in the strong sense is
indeed more complicated because chaotic dynamics manifests itself
on long time intervals.

In the examples in section V, the manifold $\M$ is not
characterized in detail. We conjecture that they are at least
dense in the unit sphere, which is most interesting to researchers
on controllability studies, but rigorous proofs have not been
found. Generally, it is a further task to investigate whether the
system is controllable on some prescribed manifolds. Providing
that the condition in Theorem \ref{ac} is satisfied, this problem
can be reduced to the transitivity of the strong ILH-Lie group
$\G$ over $S_\h$ ({\it N.B.} a group is said to be transitive if
any two points on the manifold can be connected by some
transformation in $\G$ \cite{onishchik}). While a complete list
for finite dimensional systems has given in
\cite{boothby,boothby1}, it is worth exploring the problem of
classifying all possible controllability algebras that act
transitively over the hyper-sphere. This remains to be studied in
the future.

\begin{acknowledgments}
This research was supported in part by the National Natural
Science Foundation of China under Grant Number 60433050 and
60274025. TJT would also like to acknowledge partial support from
the U.S. Army Research Office under Grant W911NF-04-1-0386.
\end{acknowledgments}

\appendix
\section{}\label{ilh}
In the viewpoint of functional analysis, the smooth domain can be
related to the Inverse Limit Hilbert (ILH-) chains \cite{omori}
defined as follows:
\begin{definition}[ILH-space]
Let $N(d)$ be the set of integers $k$ such that $k\geq d$, where
$d$ is an integer. A family of complete locally convex topological
vector (CLCTV) spaces $\{E,E^k;\,k\in N(d)\}$ is called an ILH-
chain if the following conditions are satisfied:
\begin{enumerate}
    \item Each $E^k$, $k\in N(d)$ is a Hilbert space, $E^{k+1}$ is embedded continuously in $E^k$, and the
    image is dense in $E^k$;
    \item $E=\bigcap_{k\in N(d)}E^k$, and the topology of $E$ is given by the inverse limit of $\{E^k; k\in N(d)\}$, where the inverse limit topology is the weakest topology such
    that the natural embedding $E\rightarrow E^k$ is continuous for every
    $k\in N(d)$.
\end{enumerate}
\end{definition}

For the universal enveloping algebra $E(\la)$ considered in this
paper, the smooth domain (\ref{ds}) can be equivalently expressed
as an ILH-space $\Ds=\bigcap_{k\in\mathbb{N}}{\h_k}$ by a series
of Hilbert spaces $\h=\h_0\supset\h_1\supset\cdots$ completed by
the class of scalar products \cite{nazaikinskii}
$$(\phi,\psi)_k=\langle\phi,\Delta^{k/2}\psi\rangle,\,\,k=0,1,2,\cdots,$$
where the Nelson operator $\Delta=I+L_1^2+\cdots+L_d^2$ and
$\langle\cdot,\cdot\rangle$ is the inner product of $\h$. To
properly define the series of Hilbert spaces, the Nelson operator
is required to be essential self-adjoint. With respect to this
ILH-topology, the elements in the universal enveloping algebra are
continuous and the universal enveloping algebra itself can be
exponentiated to form an infinite dimensional Frechet (or ILH-)
Lie transformation group \cite{omori,onishchik} acting on $\Ds$.
This group is no longer a Hilbert Lie group that covers most
finite dimensional Lie groups. Geometrically, the tangent space of
the group manifold at each group element is an ILH space instead
of a Hilbert space. Interested readers may consult
\cite{onishchik,omori} for more details.

\section{}\label{su11}
Let $L_\alpha=-iL'_\alpha$, $\alpha=x,y,z$. To prove this fact, it
is enough to show that every ordered $L_x^pL_y^qL_z^r$ can be
generated from the Poincare-Birkhoff-Witt Theorem
\cite{onishchik}. Let $l=p+q+r$. The case for $l=1$ is obvious.
Assume the case for $l=n-1$ is correct, we prove the validity for
$l=n$. First, if one of $L_x^pL_y^qL_z^r$ for which $p+q+r=n$ can
be generated, any other of such operators can be generated,
because we have
\begin{eqnarray*}
   & & [L_x^pL_y^qL_z^r,L_x] \\
   &=& L_x^p[L_y^q,L_x]L_z^r+L_x^pL_y^q[L_z^r,L_x] \\
   &=& -qL_x^pL_y^{q-1}L_z^{r+1}-rL_x^pL_y^{q+1}L_z^{r-1}+Q_{n-1}
\end{eqnarray*}
where $Q_{n-1}$ denotes the terms of order less than $n$.
Continuing calculating the commutations ${\rm
ad}_{L_x}^k(L_x^pL_y^qL_z^r)$, we can obtain $q+r+1$ operators in
which the orders of $L_y$ and $L_z$ range from $(q+r,0)$ to
$(0,q+r)$. Since the operators with different $(q,r)$ indices are
linearly independent, we can obtain any $L_x^pL_y^{q'}L_z^{r'}$
after proper linear combination of these commutations and
operators with order less than $n$. Similarly, we can obtain any
$L_x^{p'}L_y^qL_z^{r'}$ and $L_x^{p'}L_y^{q'}L_z^r$ from
$L_x^pL_y^qL_z^r$. Therefore any $L_x^{p'}L_y^{q'}L_z^{r'}$ can be
generated from $L_x^pL_y^qL_z^r$.

So the case of $l=n$ is valid if at least one $L_x^pL_y^qL_z^r$ of
order $n$ can be generated by lower order terms. This can be
easily verified since
$[L_x^2,L_x^{n-2}L_y]=2L_x^{n-1}L_z+Q_{n-1}$. Hence all the Lie
algebras $\A,\B,\C$ coincide with $E(su(1,1))$.

\end{document}